# Influence of step structure on preferred orientation relationships of Ag deposited on Ni(111)

Dominique Chatain[1*], Saransh Singh[2†], Blandine Courtois[1], Jérémie Silvent[3], Elodie Verzeroli[3], Farangis Ram[2‡], Gregory S. Rohrer[2], Marc De Graef[2], Paul Wynblatt[2]

[1] Aix-Marseille Univ, CNRS, CINAM, 13009 Marseille, France
[2] Department of Materials Science and Engineering, Carnegie Mellon University, Pittsburgh, PA 15213, USA
[3] ORSAY PHYSICS, ZAC Saint Charles – N95 3E Avenue, 13710 Fuveau, France

**Abstract**

Previous studies have shown that the orientation relationships which develop in hetero-epitaxy are strongly influenced by the alignment of steps in the deposit with the pre-existing steps of the substrate. In this paper we use a combination of experiments with computer simulations to identify the important influence of substrate step structure on the eventual orientation relationships that develop in the deposit. We have made use of Ag deposited on Ni as it has been used extensively as a model system for the study of hetero-epitaxy. This system displays a large lattice mismatch of 16%. It is shown that on any surface vicinal to Ni(111), which has two possible kinds of <110> steps (A-steps with {100} ledges and B-steps with {111} ledges), a Ag deposit adopts a single orientation relationship because only A-steps remain stable in the presence of Ag.

---

[*] corresponding author
[†] currently at Lawrence Livermore National Laboratory, Livermore, CA 94568, USA
[‡] currently at EVRAZ NA, Portland, OR 97203, USA

# 1. Introduction

There is a strong interest in improving the understanding of the fundamentals of interface formation in metal-substrate heteroepitaxy. This is indeed a key-issue for thin films in which physical properties and morphological stability depend on the microstructure. Thin solid metal films prepared by physical vapor deposition onto a substrate are mainly polycrystalline. When the substrate is a single-crystal, the grain boundaries (GBs) in the deposited or the annealed film may be reduced, or the GBs may be completely eliminated by the development of grains which have a single low-energy orientation relationship (OR) with the substrate. Identifying the physical parameters which drive the growth of the grains of a single variant of an OR would allow the production of single-crystalline films, thereby improving the durability of devices. This issue is addressed in this paper by using a combination of experiments and modeling on the Ag-on-Ni model system.

Atomically flat single-crystal substrates are appealing, and are frequently employed to grow thin films, because they tend to minimize interfacial defects. They usually have high crystallographic symmetry (such as the {111} and {100} surfaces of face centered cubic (FCC) metals, silicon and germanium, or the c-plane of sapphire) that stabilizes several orientation variants of the grains of a film, which thus remains polycrystalline (see for example [1,2]). However, substrates larger than a few microns are never perfectly flat at the atomic scale, so that even in the field of microelectronics, silicon wafers, for example, are slightly miscut and curved such that their surface consists of large atomically flat terraces of the densest nominal surface plane, separated by steps that may be either one or several atomic units in height. There is an advantage to the presence of these steps, because, by breaking the high symmetry of the substrate surface, they favor one of the grain orientation variants, and the growth of single crystal thin films [3]. Therefore, controlling the nature of the substrate step-structure, and identifying the types of substrate miscut orientations, and step directions, that will produce single-crystalline films, may lead to improvement of the film stability by limiting the presence of GBs.

This paper presents a study of the OR adopted by Ag on Ni(111), which takes advantage of the extensive body of literature on the heteroepitaxy of Ag films on different Ni substrate orientations, acquired by both experiments and atomistic simulations [4-21]. The lattice parameters of Ni and Ag are 0.352 nm and 0.409 nm, respectively; this corresponds to a 16% lattice mismatch. Our own previous work [4,5] has shown that Ag films adopt a unique (preferred) orientation relationship that depends on the Ni substrate orientation, and that is strongly influenced by an alignment of the interfacial steps in the Ag deposit with the natural steps present on the substrate. On Ni(111) and (100) substrates, several variants of the Ag grains are most stable when their {111} planes are parallel to either of these two Ni substrate surfaces, and at least one of the <110> directions of the two abutting Ag and Ni interfacial planes are parallel. Thus, when these planes are strictly flat, symmetry allows four Ag variants on Ni(100) (two plus their twin-related variants), and two (twin-related variants) on Ni(111). On Ni(111), the Ag films can adopt either a cube-on-cube orientation relationship (*OR C*) or a hetero-twin orientation relationship (*OR T*) also named normal and reverse, respectively (see for example [6]). The previous experimental determination of the ORs of Ag equilibrated on more than 200 different Ni substrate surfaces [5] has shown that *OR C* prevails on Ni substrates that have orientations which correspond to miscuts from (111) towards another adjacent {111} plane, whereas when Ni(111) substrates are oriented so as to correspond to miscuts towards a {100} plane, the OR of Ag on Ni departs from *OR T* by a small rotation angle about the common <110> in-plane direction along the step edges [4,5,7]. It should be noted that when a FCC (111) substrate is miscut towards another adjacent {111} surface the resulting step ledges will display a {111} structure; in contrast, when a FCC (111) substrate is miscut towards a {100} plane the step ledges will display a {100} structure. This is shown in Fig. 1a. In previous literature (see for example [22,23]) the {100} and {111} step ledges at the edge of a {111} terrace have been referred to as A- and B-type steps, respectively, as illustrated in Fig. 1b. It is worth mentioning that the ledges of A- and B-steps have also previously been refered to as {111} or {100} "microfacets" by van Hove and Somorjai [24], or that A- and B-steps have been described as <110>/{111} and <110>/{100} steps by Michely and Comsa [25]. In what follows we use the terms A- and B-steps to distinguish them from each other.

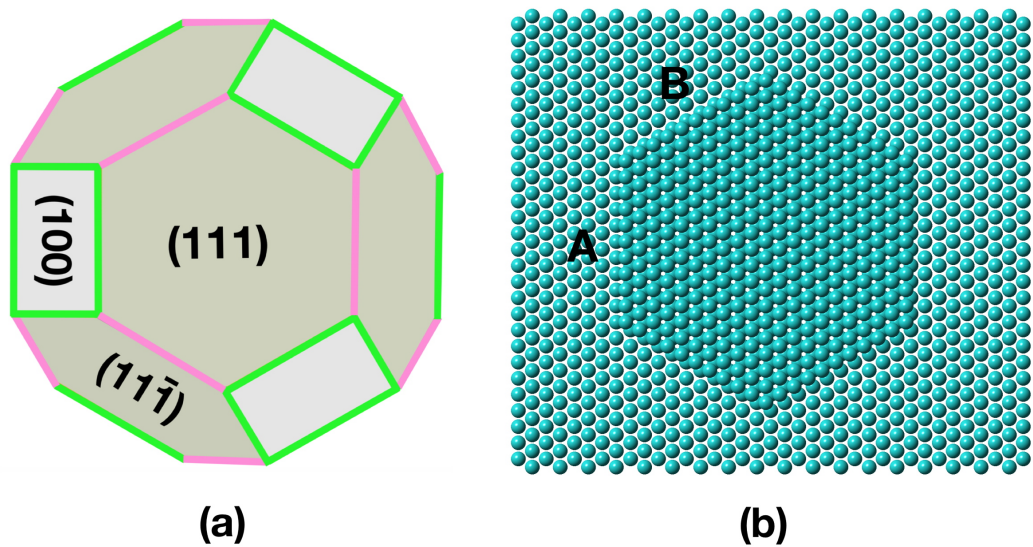

*Figure 1. (a) shows the relationships between a {111} plane and the adjacent {111} and {100} planes, and (b) illustrates the atomic structure of a hexagonal monolayer on top of a {111} FCC surface; the monolayer displays alternate A-type and B-type steps with ledges oriented along {100} and {111} surfaces, respectively (and shown in green and pink in (a)).*

The epitaxial growth of a few atomic layers of Ag on Ni(111) substrates has been addressed in several papers, with a focus on the structure of the first deposited layer [6,8]. It is worth noting that the difference in OR between *OR C* and *OR T* can only be determined if at least two complete atomic layers of Ag are deposited. It has been shown in [6] that when Ag is deposited under vacuum, very thin grains of the two ORs (*OR C* and *OR T*) coexist but their relative frequency depends on the deposition temperature. Up to room temperature there is an equal proportion of the two ORs, but at 400 K, the grains with *OR T* occupy most of the surface. A recent paper which has studied the orientation of Ag deposited on the vicinal Ni(11 9 9) substrate [21], that corresponds to a miscut towards (100), has shown that Ag grains grow in *OR T*, and is thus in agreement with our previous results [5,7]. To the best of our knowledge there have been no previous studies of the OR of Ag on a surface vicinal to Ni(111) with an orientation corresponding to a miscut towards {111}, except for observations in our own previous published work [5] of Ag on Ni(17 17 10), which deviates from (111) by 12°.

This paper addresses the relative stabilities of *OR C* and *OR T* for Ag on Ni substrates with orientations vicinal to (111). We have investigated the microstructure evolution of Ag polycrystalline films that occurs by annealing between 373 K and 873 K. The morphology of the film has been tracked *in-situ* in an ultra-high-vacuum scanning electron microscope (UHV-SEM) and the ORs have been analyzed *ex-situ* using electron backscattered diffraction (EBSD), as in [5]. We show that Ag grains with the hetero-twin (or reverse) OR may be produced by annealing even on surfaces with step ledges of the {111} orientation and that the growth of the *OR T* grains takes place at the expense of the *OR C* grains. We have also performed atomistic simulations in order to investigate the evolution of step structure and how it leads to a preference for *OR T*.

## 2. Material and methods

The substrate used consisted of a polycrystalline <111> textured Ni film, with grains larger than 100 µm, grown on the ($1\bar{1}02$) r-plane of sapphire. The details of sample preparation have been previously described in [3]. Briefly, the <111> textured Ni films were obtained by annealing 560 nm thick Ni films at 1373 K, in one atmosphere of an Ar - 40% $H_2$ gas flow. Compared to the usual preparation of Ni(111) substrate surfaces by polishing, as in [6], our method produces large areas of clean Ni(111) surfaces without miscut. Figure 2 displays AFM images of the (111) surfaces of the Ni film grains. Figure 2a shows almost flat (111) surfaces with terraces several microns in width, delineated by approximately circular edges corresponding to 3±1 nm high steps. A higher magnification view of an upper terrace in Fig. 2b shows a network of 3 sets of straight steps running in directions lying at 120° to each other. These steps form during sample cooling to relieve the tensile stress in the Ni film, which results from a larger thermal expansion coefficient of nickel than that of sapphire. During this deformation, dislocations in the bulk of the Ni film undergo glide on the three {111} planes tilted by 70.5° to the (111) surface plane, under the constraints of thermal stresses, and emerge at the surface along the three equivalent <110> directions of the (111) plane, forming steps with {111} ledges (i.e., B-type steps). The deviation from a flat (111) surface is from 1 to 3 times the spacing between Ni(111) atomic layers, (i.e., ~ 0.21 nm each). Just prior to Ag deposition, the samples were reheated under Ar - 40% $H_2$ flow to between 1073 K and 1423 K for one hour in order to reduce any NiO surface layer formed by previous exposure to air. These conditions ensure oxide reduction and the removal of oxygen as water vapor [26,27].

After the above preparation procedure, the polycrystalline Ni(111) substrates were rapidly transferred through air to the metal deposition vacuum chamber, in which Ag thin films were deposited at room temperature at a rate of 0.7 nm/sec under a vacuum of $1.6 \times 10^{-4}$ Pa, from melted pellets of 99.99% pure Ag (supplied by the Kurt J. Lesker Company) contained in a tungsten crucible. Films of either 20 or 30 nm in thickness were prepared. Any oxygen adsorbed on the Ni surface during sample transfer was further removed from samples after Ag film deposition, during annealing performed either under a flowing Ar - 40% $H_2$ atmosphere at a rate of 10 cc/min, or under UHV ($10^{-8}$ Pa base pressure) in the *in-situ* UHV-SEM. In the latter case, oxygen dissolves into the bulk of the Ag and Ni films during annealing at temperatures in the range of 473 K to 873 K [26-29].

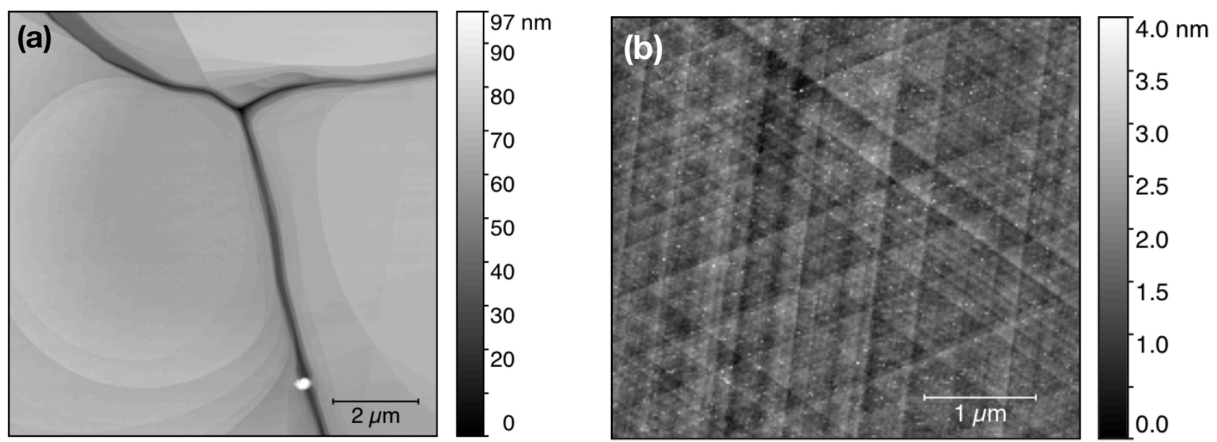

*Figure 2:* AFM images (a) of a 10x10 µm Ni surface at the junction of 3 grains showing several µm-wide terraces with almost circular macro steps 3±1 nm high; (b) details of a 4x4.3 µm region of the flat Ni(111) terrace of (a) which shows a network of straight steps running along 3 directions at 120° to each other. The step height here is 1 to 3 times the spacing between atomic Ni(111) planes (i.e. 0.21 ±0.05 nm each).

The ORs of the grains of the Ag films on the underlying Ni(111) (textured poly-crystalline substrate) were determined by automated EBSD mapping in a SEM after annealing. The orientations of Ag and Ni were measured simultaneously during a single experimental run. This ensures an accuracy of the relative orientation between the Ag and Ni crystals of better than 0.3°. The ORs of the Ag grains on each Ni(111) grain were determined by comparing the pole figures (PFs) and/or the stereograms of the two phases obtained from the EBSD data, as described in a previous paper [5]. Two microscopes were employed in this study. One was located at ORSAY-PHYSICS in Fuveau, France; this is a LYRA3 from TESCAN, equipped with a field emission gun and a combined electron dispersive x-ray spectroscopy (EDS) with an EBSD system from OXFORD Instruments, controlled by the AztecHKL software package. The other microscope was located at Carnegie Mellon University (CMU) in Pittsburgh, PA, USA; this is a FEI Helios plasma focused ion beam SEM equipped with an Oxford-Aztec orientation imaging system and the Symmetry EBSD detector. Data were acquired at 20 or 30 kV with the substrate plane tilted by 70° with respect to the horizontal plane. The EBSD data obtained from both cameras were post-processed with the TSL/EDAX OIM software package.

**3. Experimental results**

As mentioned above, annealing was performed either under an Ar-$H_2$ atmosphere or *in-situ* under UHV at temperatures ranging from 473 to 873 K. In all cases, the Ag grains on the surface of each Ni(111) substrate grain displayed the same ORs, but the Ag film ended up in two different morphologies depending on the conditions of the annealing treatment and its initial thickness: thicker (30 nm) films maintained their film-substrate configuration whereas thinner (20 nm) films underwent dewetting which eventually led to break-up of the film into islands.

*3.1. Annealing of 30 nm films, in-situ under UHV*

The consequences of annealing 30 nm thick films was observed *in-situ* in the UHV SEM. Heating was performed at temperatures up to 873 K, in steps of 100 K. Images were acquired after cooling down to 473 K in order to prevent Ag condensation in the electron column.

Figure 3 presents a set of images which illustrate the evolution of a Ag film upon successive annealing treatments. It shows that Ag film break-up proceeds by the nucleation and growth of holes at Ni surface defects. Holes in the Ag film nucleate first at Ni grain boundary grooves and triple junction pits, and then along the macro-step edges present on each Ni(111) crystal. The holes appear as darker regions in each image and some of them are identified by arrows. Upon further annealing, holes in the Ag film grow larger, and deviate from circular shape because of either Ag surface energy anisotropy or as a result of hole coalescence (see Figs. 3b, 3c, 3d). There is no visible edge hump characteristic of dewetting films [30] surrounding the holes, neither at this scale of observation nor at a 10 times larger magnification, as was observed on the sample fully broken-up into discrete particles under a hydrogenated atmosphere (see Figs. 4). This is presumably related to the very rapid surface self-diffusion on the (111) Ag surface as explained in [31]. Auger analysis of the film surface chemistry after annealing did not show any sign of oxygen, but did indicate the presence of a trace of sulfur.

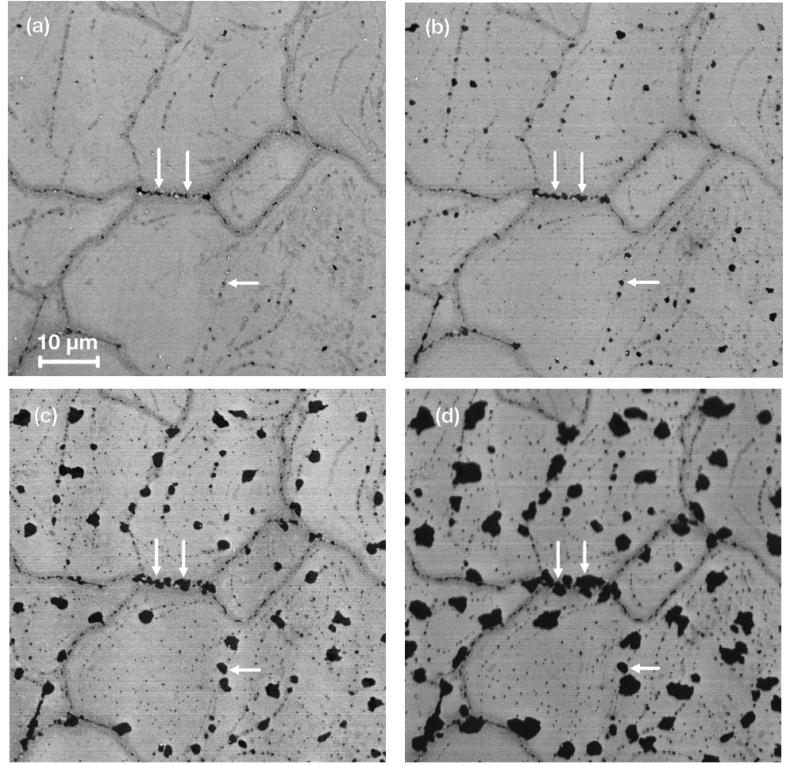

*Figure 3*: Evolution of a 30 nm thick Ag film upon successive anneals on a <111> textured polycrystalline Ni substrate; (a) after a first anneal for 1 h at 733 K, (b to d) after additional anneals at 873 K (b) for 1 h, (c) for 2 h, and (d) for 4 h. The darker contrast in each image corresponds to holes in the Ag film which have exposed the Ni substrate. The Ni grain boundaries underneath the Ag film are clearly visible (medium grey). The other darker "dashed" lines within the Ag grains correspond to underlying Ni macro-steps similar to those shown in the AFM image of Fig. 2a, and on which holes have nucleated in the Ag film. Vertical and horizontal arrows indicate holes nucleated at Ni grain boundaries and at macro-steps, respectively. In (b), (c) and (d) the arrows point to the same holes after successive anneals.

Figure 4 shows back scattered electron images and processed EBSD data, acquired in the SEM-EBSD of ORSAY-PHYSICS, of a region similar to that of the sample displayed previously in Fig. 3d, after the last annealing step at 873 K. Fig. 4a is a topographic image assembled by using forward scattered electron images taken from different directions, showing the GBs between several Ni grains, as well as the holes in the Ag film on each of the Ni grains. Small convex Ag crystals with the brightest contrast have formed along the Ni GBs where the Ag film is prone to dewet because of the lack of flatness of the Ni substrate in these regions. Fig. 4b is an inverse pole figure map in the Z direction (Z-IPF), perpendicular to the sample surface, of the region displayed in Fig. 4a. The map is uniformly blue, with the exception of a small pink needle which is a (511) Ni-twin of an adjacent Ni(111) grain. In Fig.

4b, most of the Ni GBs are marked by white dashed lines. The uniform blue color indicates that both the Ag film, as well as the Ni surfaces seen through the holes in the Ag film, have a (111) pole perpendicular to the surface, according to the color code indicated in the standard stereographic triangle (SST) of the inset in Fig. 4b.

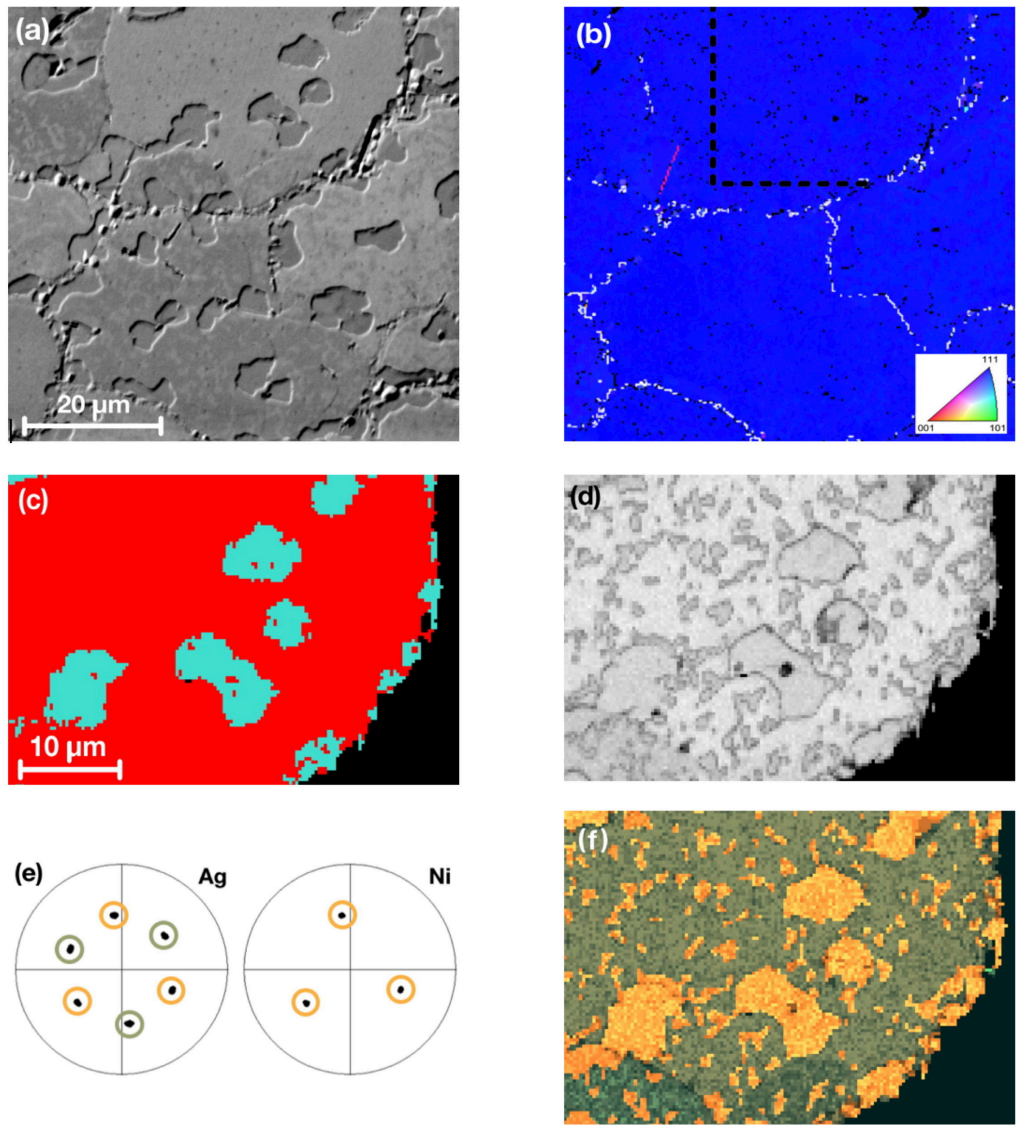

*Figure 4: Back scattered electron image and EBSD data maps of a 30 nm thick Ag film on a <111> textured Ni polycrystal, after the final treatment applied in Fig. 3d. (a) Topographic image showing the holes in the Ag film on each of the Ni grains, and the GBs between several Ni grains outlined by small convex Ag crystals broken-up from the original continuous deposited film. (b) Z-IPF map showing that the <111> z-axes of the Ag film and of the exposed Ni in the dewetted surface are parallel. (c) Chemical map on a single Ni grain where Ag is in red and exposed Ni in the holes of the Ag film in light blue. (d) Band contrast map which shows that the Ag film is made up of a single-crystal containing several small island-grains; (e) {100} pole figures of the Ag grains in Fig. 4d, showing the presence of two Ag orientations (orange and green) and only one orientation (orange) for Ni in the holes of the Ag film. (f) image quality map superimposed on an orientation map of the Ni and the two types of grains in the Ag film colored according to the color scheme used in the {100} pole figures of Ag and Ni in Fig. 4e. The small Ag island grains in the film are in cube-on-cube OR (OR C) with the Ni grain underneath while the large Ag grain is in hetero-twin OR (OR T); i.e. the Ag film is bicrystalline.*

Figures 4c to 4e provide a detailed analysis of the Ag film microstructure on a single Ni grain identified at the top right of Fig. 4b by a Ni GB on the right and two black dashed lines. Fig. 4c is a chemical map of the selected region taken by energy dispersive x-ray spectroscopy (EDS) where Ag is colored red and Ni colored light blue. Ni is detected in the regions where holes have formed in the Ag film, and at the GB edge, on the right of the image. Fig. 4d is a band contrast image of the EBSD patterns acquired in the selected region; the darkest contrast corresponds to a poorer indexation, due to the overlap of two EBSD patterns at a GB. It shows that the Ag film, which is continuously red in Fig. 4c, contains numerous small grains of a different orientation, even though both Ag orientations must have a <111> normal. According to the {100} pole figures of Ag and Ni shown in Fig. 4e, there are two different in-plane orientations for Ag, one of which (indicated by orange circles) is identical to that of the Ni grain, and another (indicated by green circles), which is in a twin-related orientation rotated by 60° from the orange orientations. The two orange and green colors used in the {100} PF have been assigned to each region of these two orientations and superimposed on an image quality map in Fig. 4f. It clearly shows that on a given Ni(111) grain the Ag film consists of a large region oriented in *OR T* (green), while the small Ag island-grains are oriented in *OR C* (orange). These results, presented for one of the grains of the Ni(111) substrate, apply to all the other Ni(111) grains, each of which behaves as an independent Ni(111) single-crystal substrate. The proportion of Ag islands within the large Ag crystals measured on several of the Ni(111) grains is 26±4%. Thus, *OR T* is the preferred OR for Ag on Ni(111) after 4 h of annealing at 873 K. It is worth noting that the GBs within the Ag film between the main continuous grains and its islands are $\Sigma 3$ boundaries.

*3.2. Annealing of a 20 nm thin film at 873 K in Ar-$H_2$*

A thinner 20 nm film was annealed in a furnace under a flow of Ar-$H_2$ in two steps: for 1 h at 673 K followed by 1 h at 873 K. The rate at which the sample reaches the set-point temperature is twice as fast as in the *in-situ* SEM. Because of the lower film thickness, this annealing treatment was sufficient to break the Ag film up into islands.

Figure 5 displays secondary electron images together with EBSD data, acquired in the SEM-EBSD at CMU, from a sample in which the Ag film has broken-up into discrete Ag islands on the surfaces of the <111> textured Ni grains. As for the sample investigated in the previous section, each Ni grain acts as a different Ni(111) substrate on which all the Ag islands behave identically. Figs. 5a and 5b show that the Ag islands adopt two different shapes; they are either flat, irregular pieces of film a few microns long and less than a micron wide, or convex faceted grains with a contact diameter of less than one micron, many of which display 3-fold symmetry. Figs. 5c and 5d display IPF maps of Ni and Ag, respectively. The chemical nature of the two phases has been identified using EBSD. Since the traditional Hough based method is insensitive to similar crystal structures with different lattice parameters, such as FCC Ag and FCC Ni, this analysis was performed using the Dictionary Indexing (DI) method [32]. The experimentally obtained backscatter diffraction patterns were compared with simulated patterns for Ag and Ni using the dot product similarity metric. The differences in intensity along the Kikuchi bands are sufficient for the DI approach to differentiate between the phases (the confidence index values reveal that the differences between the two phases are well above the variation of the top dot product within each individual phase). An open source implementation of this method can be found in [33].

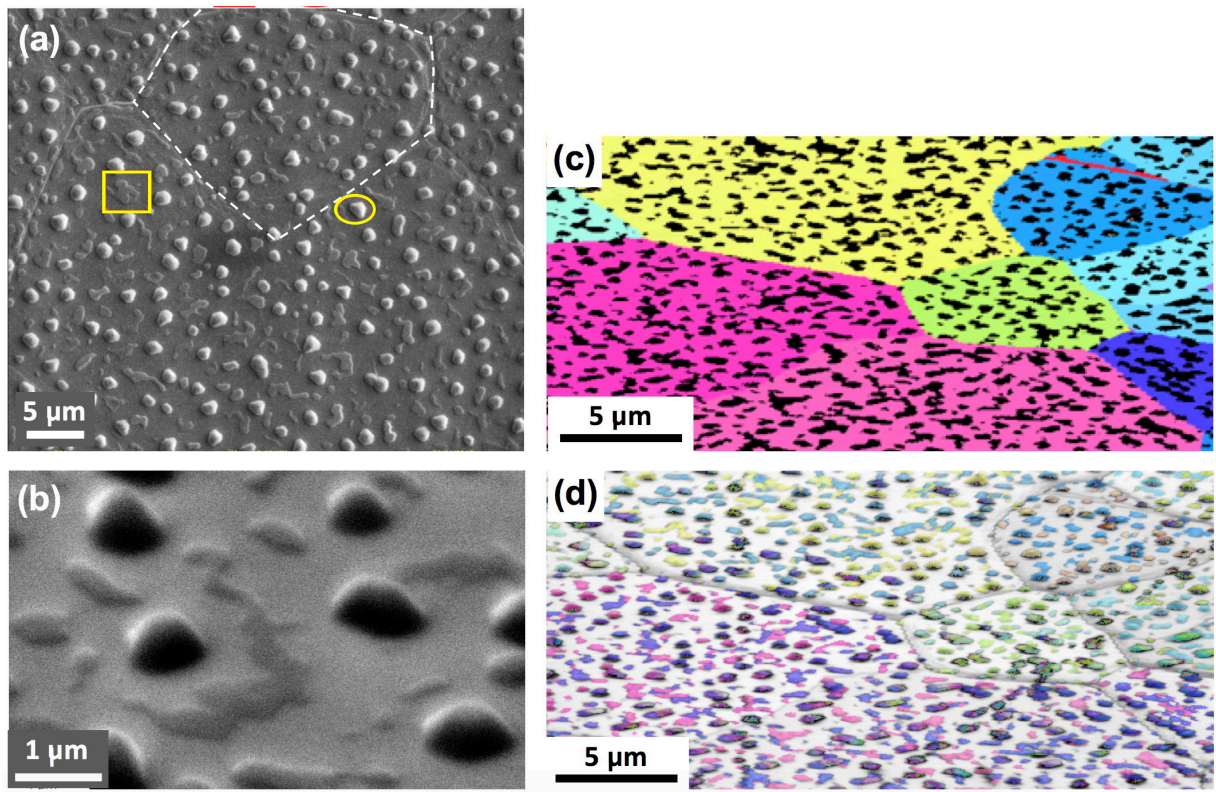

*Figure 5: (a) Secondary electron image of a top view of a dewetted 20 nm film after annealing under a hydrogenated atmosphere. The yellow square and circle surround flat and convex Ag islands, respectively. An example of a Ni grain boundary is marked by a white dotted line. (b) Magnified region of the sample tilted by 70°, showing the convex and flat Ag islands in a single Ni grain. (c) and (d) IPF maps of Ni and Ag respectively, in which colors have been assigned to the Ni grains and their Ag particles according to their in-plane directions, such that the two ORs on each Ni grain can be identified. The unidentifed black regions in (c) correspond to the Ag grains shown in (d). A band contrast map has been added to the Ag IPF map in (d) to emphasize the Ag particle shapes and the location of the Ni GBs. Note that there is no systematic difference between the ORs of the flat and the convex Ag islands.*

From images such as Fig. 5b, the contact angle of the convex Ag crystals which have reached their equilibrium shape on the Ni substrate was determined from their height (h) and their contact diameter ($2r°$) while the sample was tilted by 70° for EBSD analysis. By approximating particle shape as a spherical cap, the contact angle $\theta$ may be obtained from $\tan(\theta/2) = h/r°$. The height of the particle h equals $H\cos(70°)$, where H is the height of the tilted particle above the contact diameter. An average value of θ = 29±3° was obtained from measurements on 5 particles. This result may be compared to the measurements of Nagesh and Pask [34] who report small and large contact angles of 9° and 90° for liquid Ag on Ni and NiO, respectively, just above the melting point of Ag. From a comparison with those results, we conclude that there was no oxide at the surface of Ni in our samples.

The flat irregularly shaped Ag islands, produced during dewetting of the film have not yet reached the near-equilibrium shape developed by the convex islands. As explained in [35], a single-crystalline piece of Ag film will reach its Wulff shape if defects such as dislocations (with screw components perpendicular to the surface) are present to provide nucleation sites that might foster the addition of new surface atom layers and lead to 3-D growth. The presence of both flat and convex islands is an indication that the flat particles presumably lack suitable dislocations for 3-D growth. As seen for the thicker film, the flat Ag particles do not display a dewetting hump at their periphery presumably because of a very rapid Ag surface self-diffusion ($5 \times 10^{-7}$ cm$^2$/sec at 873 K for the (321) under hydrogen [36] and probably faster on the (111) as shown by calculations on Cu [37]).

In Figs. 5c and 5d it can be seen, for example, that on the pink Ni grain (bottom left of Fig. 5c), some of the Ag islands (Fig. 5d) are also pink, while others are blue. Using the (111) PF of each Ag grain, and of the Ni substrate (not shown here), it is observed that most of the Ag islands are single-crystals with their (111) planes lying parallel to the Ni(111) surface, and an in-plane orientation which corresponds to either *OR C* or *OR T* with respect to the Ni(111) grain. A few islands are twinned bi- or tri-crystals with Σ3 grain boundaries (i.e. made up of crystals having alternating *OR T* and *OR C* with respect to the substrate). The same behavior is displayed by the Ag islands on each of the other Ni grains of Fig. 5c. Comparing secondary electron images and EBSD maps such as those of Fig. 5c and 5d, it is observed that the in-plane orientations of the Ag islands on Ni are not correlated to their shapes (flat or convex). A quantitative analysis of almost 300 islands from different Ni(111) grains shows that *OR C* and *OR T* both occur with about the same frequency.

**4. Atomistic simulations and interfacial steps**

In previous studies of the ORs of Ag films on Ni, we have proposed important roles for Ni steps (in the context of the terrace-step-kink description of surface structure) in the development of the equilibrium ORs. Current experimental techniques do not allow direct observations of atomic size steps at buried interfaces, whereas atomistic simulations do allow a means of investigating possible interface structures at the atomic scale.

Atomistic simulations have been performed on Ni substrates vicinal to Ni(111): Ni(11 9 9) substrates, lying about 5.6° from (111), with A-type steps, and Ni(11 11 9) substrates, lying about 5.2° from (111), with B-type steps. Recall that A-type steps tend to stabilize *OR T*, whereas B-type steps tend to stabilize *OR C*.

The simulation procedures have been described previously [4,38]. Here we provide a brief overview. Modeling was conducted by molecular dynamics (MD) and molecular statics (MS) simulations, employing the LAMMPS code [39,40] in conjunction with embedded atom method (EAM) potentials [41]. These potentials have been used extensively in previous modeling studies of the Ni-Ag system (see [4] and references therein). The computation cells used consisted typically of ~16,000 Ni atoms and either 0 or ~10,000 Ag atoms. The MD equilibration portion of the simulations was applied for time intervals of 12 and 16 ns at a temperature of 900 K. Since no significant changes took place between these two equilibration times, only the 16 ns equilibrations are shown below. When Ag was added to the Ni substrates, the Ag atoms were initially distributed in random slabs placed in contact with the Ni substrates, so as to ensure that any OR between the Ag and the Ni substrate, resulting from equilibration, was unbiased. Computation cells made use of periodic boundary conditions in the x- and y-directions, and were terminated by free surfaces in the z-direction perpendicular to the substrate surface.

4.1 Results of computations

We first consider the results obtained on Ni substrates without any added Ag. The results of equilibrating the substrates for 16 ns at 900 K are shown in Figs. 6(a) and (b) for the Ni(11 9 9) and Ni(11 11 9) substrates, respectively. Initially, both of these substrates had 4 steps each and 40 atoms along each step. The equilibration at 900 K has caused a slight step roughening, with both substrates having essentially equal numbers of atoms displaced from their step positions. This result is consistent with the experimental findings on surfaces with the A- and B-type steps of other FCC metals, such as Pt [42,43], Ag, Cu [44] and Pb [45], that the B-step (with a {111} ledge) has a slightly lower (or equal) energy and a somewhat larger (or equal) stiffness than the A-step (with a {100} ledge).

We now compare these results with the same two Ni substrates equilibrated in contact with Ag films for 16 ns at 900 K. The results obtained are displayed in Figs. 6c and 6d after removal of the Ag film, so as to reveal the structure of the Ni substrate interface. There is now a striking difference between the behaviors of the Ni(11 9 9) substrate with A-type steps (Fig. 6c) compared with the Ni(11 11 9) substrate with B-type steps (Fig. 6d). Figure 6c displays essentially the same degree of step roughening as was obtained by equilibrating without Ag on the surface, as shown in Fig. 6a, whereas Fig. 6d has now undergone quite significant roughening. Clearly, the presence of a Ag film in contact with the B-type steps of the Ni(11 11 9) substrate has decreased step stability and increased the mobility of the Ni atoms at the interface.

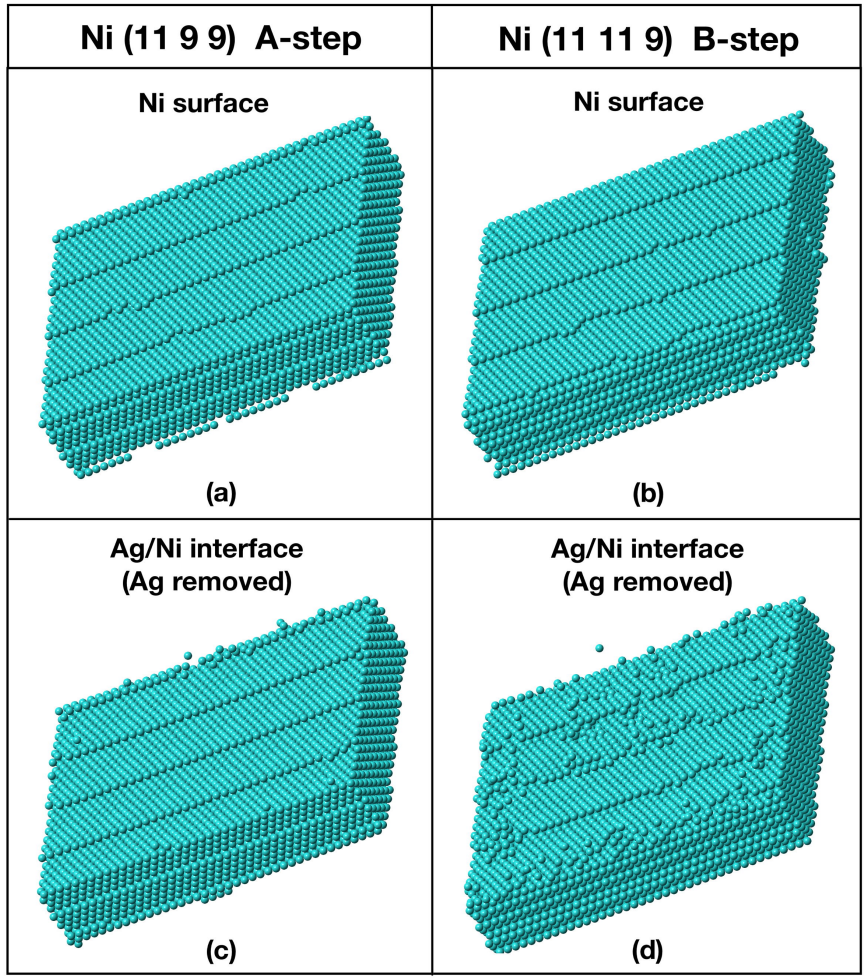

*Figure 6:* Stability of steps in the absence and presence of Ag on substrates vicinal to Ni(111): Ni(11 9 9) with A-type steps, and Ni(11 11 9) with B-type steps, all equilibrated for 16 ns at 900 K. (a) Ni(11 9 9) substrate and (b) Ni(11 11 9) substrate. Neither (a) nor (b) had any Ag films on their surfaces. (c) Ni(11 9 9) substrate, and (d) Ni(11 11 9) substrate. Both (c) and (d) were equilibrated with Ag films about 12 monolayers thick. In (c) and (d) the Ag films have been removed so as to reveal the Ni step structures.

It is also possible to extract the ORs displayed by the Ag films on the Ni substrates from the simulations. This can be done by constructing a pole figure from the computed atom positions, as described in a previous paper [4]. Here we provide a brief description for the sake of completeness. Consider a FCC crystal constructed in a coordinate system with the z-axis perpendicular to the (hkl) surface (x-y plane) orientation. Since the 12 nearest neighbors of any atom will lie along <110> directions, the vectors which correspond to the nearest neighbor directions in the sample coordinate system will yield a <110> PF when projected onto a stereogram. The orientations of Ag and Ni in a simulation may therefore be deduced by computing the average directions of nearest neighbors of all atoms and displaying them in a stereographic projection.

The results of this type of analysis are shown in Figs. 7a and 7b in the form of <110> PFs produced from the simulations of Ag films on Ni(11 9 9) and Ni(11 11 9) substrates, respectively. The Ag<110> poles are shown by red lozenge symbols and the Ni<110> poles are shown as smaller blue lozenges. In these figures we also show the Ni[111] pole (as a green triangle) and the Ni[100] (as a green square). It is important to note that whereas there are only 3 Ni<110> poles surrounding the Ni[111] pole, as expected for a FCC structure, there are 6 Ag<110> poles surrounding the Ni[111] pole, 3 of which overlap the Ni<110> poles and 3 others which are rotated by 60° with respect to the Ni<110> poles that surround Ni[111]. The Ag poles that overlap the 3 Ni poles originate from atoms that belong to *OR C* Ag grains, and the Ag poles that are rotated by 60° from those 3 Ni poles originate from atoms that belong to *OR T* grains. Thus it is possible to evaluate the relative numbers of Ag atoms with surroundings that place them in *OR C* or *OR T* grains.

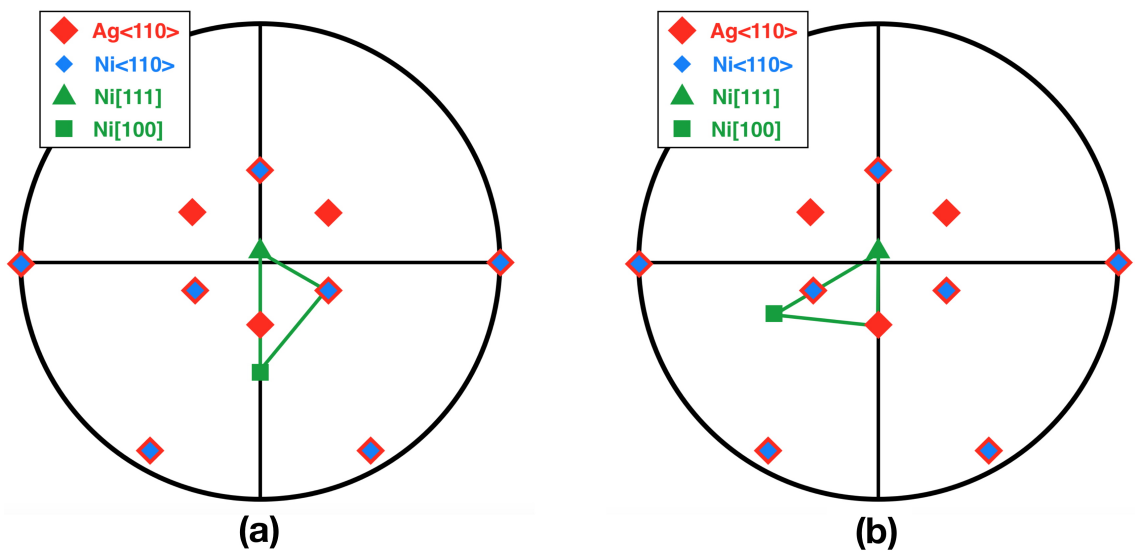

*Figure 7: Pole figures of Ag films (a) on a Ni(11 9 9) substrate and (b) on a Ni(11 11 9) substrate. The red and blue symbols represent the <110> poles of Ag and Ni, respectively. In addition, the Ni [111] and [100] poles are displayed with green triangle and square symbols, respectively, and the Ni standard stereographic triangle is shown by green lines, for reference, (Color on line).*

One complicating factor in this type of analysis is the presence of stacking faults in the Ag film. These defects often originate from Ni steps at the Ni-Ag interface. Stacking faults are characterized by two (111) atom layers which have nearest neighbor surroundings that correspond to a hexagonal close packed (HCP) coordination. If the nearest neighbors of an atom in HCP coordination are interpreted as having FCC coordination, they would lead to spurious <110> poles. Thus, when stacking faults are present, any Ag atoms with HCP coordination must be omitted from consideration in the PF analysis. Fortunately, it is easy to distinguish between atoms that have HCP or FCC coordination, by means of the common neighbor analysis (CNA) [46]. Thus, CNA was initially applied to all Ag atoms in the simulations, and any Ag atoms with HCP coordination were not used in the construction of the PFs of Figs. 7. From the numbers of Ag atoms that contributed to the <110> poles of Fig. 7a for the Ni(11 9 9) substrate, with A-type steps, we find that 87% of Ag atoms were in grains with *OR T*, whereas in Fig. 7b for the Ni(11 11 9) substrate, with B-type steps, 99% of Ag atoms originated from grains in *OR C*.

## 5. Discussion

The present experiments show that the thicker Ag films, which remain polycrystalline, also retain a {111} interface orientation parallel to the {111} Ni substrate surface. However, the Ag grains adopt two different ORs on any given Ni grain: *OR T* and *OR C*, with *OR T* being dominant. The computer simulations show that the presence of A-type steps on Ni favor the development of *OR T*, and that B-type steps favor *OR C*. The Ni substrates are not atomically smooth, and contain different kinds of steps, as shown in Fig. 2. The steps in Fig. 2a are circular, and therefore must display segments of both step types. The Ni surfaces also contain smaller steps, that are all of B-type, as displayed in Fig. 2b. However, as illustrated by the results of the computer simulations (in Fig. 6d) B-type steps are not stable at the Ni-Ag interface, and tend to roughen so as to display multiple different step types. Thus, it is not surprising that the Ag films generally contain both types of OR.

Thinner Ag films undergo dewetting to form discrete islands. Most of the islands are single crystals, and again both *OR C* and *OR T* islands are observed to coexist. Although one of these two ORs possibly leads to lower energy islands than the other, once dewetted islands are formed, it is unlikely for the less stable islands to rotate into the other OR because of intermediate high energy configurations.

In the experiments, the initial Ag films contain grains or islands of the two ORs. When the Ag film is thicker, the Ag grains in *OR T* grow at the expense of the grains in *OR C* to produce a film which is composed of large *OR T* grains containing small *OR C* island grains, the area of which covers only a small fraction of the film. When the Ag films are thinner they tend to break up rapidly, before much grain growth has occurred, so that the proportion of *OR C* grains remains relatively higher.

The simulations show two features: (i) the Ni A-steps stabilize *OR T* while the B-steps stabilize *OR C*; and (ii) when Ni substrates are in contact with Ag, Ni B-steps are destabilized and roughen into segments containing A-steps, whereas the A-steps remain straight. There are no specific data in the literature on the effects of step ledge structure (i.e., whether steps are of type A or type B) on the resulting orientation relationship between abutting phases.

The main differences between calculations and experiments lie in the time-scale and the sample size. In calculations, the Ag film in contact with a Ni surface that contains a given type of steps, crystallizes very quickly with an OR corresponding to the initial type of steps present. While Ag grains of the two ORs coexist, the calculation time is too short for neither much grain growth to occur, nor for the ORs to change in response to the modification of the Ni substrate step morphology. What we learn from the calculations is that when Ni is in contact with Ag, the step morphology evolves in a manner which privileges *OR T*. In the experiments, the time scale is long enough for grain growth to produce an increase in *OR T* grains at the expense of the grains in *OR C*.

Previous studies of Ag on Cu(111) [47] and of Pb on Ni(111) [48] systems, which are comparable to Ag on Ni(111) in the sense that they have a large difference in atomic size between the deposit and the substrate, also report the same preferred stability for *OR T* over *OR C*. There is little chance that the substrates in these studies were perfectly flat or that they were systematically miscut in a direction that favored *OR T*. It is worth mentioning a recent study on the stability of Pt steps at vicinal (553) and (533) surfaces [49]. The Pt steps (one atomic unit high) on top of Pt (111), are of A- and B-type for the (533) and (553) surfaces, respectively. Upon hydrogen adsorption, the B-type step "roughens" into A-type segments. Calculations have shown that this is related to a strong decrease of the B-step energy compared to the A-step energy due to hydrogen adsorption. Here, our observations coupled with calculations suggest a similar kind of behavior at an interface, rather than at a free surface; in our case, this step reconstruction also controls the OR of the crystal that grows on the (111) substrate.

The findings of this paper emphasize the critical role of interfacial steps and their structures in controlling the OR, rather than the conventional wisdom that the OR is determined by the energy of an ideal atomically flat interface.

## 6. Conclusions

In our previous work [4,5,7] we have demonstrated the important role that is played by substrate steps on the resulting OR adopted by the deposit. This influence manifests itself as an alignment of the steps of the deposit with the pre-existing steps of the substrate. In this paper we have also shown that when a substrate surface can display steps with differing structures, these can produce different deposit ORs. In particular, the Ni(111) substrate investigated here can present two possible step structures, i.e., A- and B-steps. Experiments performed on Ag deposited on Ni(111) surfaces, which contain both types of steps, show that the initial Ag deposit consists of two ORs, *OR T* and *OR C* in approximately equal proportions. When such samples are equilibrated by annealing for longer times, the proportion of grains in *OR T* tends to increase.

It has been possible to interpret these results by performing MD simulations on two different Ni substrates, namely Ni(11 9 9), with a surface that consists of {111} terraces separated by A-type steps, and Ni(11 9 9) which consists of {111} terraces separated by B-type steps. When both of these bare Ni substrate surfaces (i.e. without Ag) are simulated, the two different step structures are about equally stable. However, when the simulations are performed with Ag on the surface, the B-steps on the Ni(11 11 9) substrate are considerably roughened indicating destabilization, whereas the A-steps on Ni(11 9 9) are essentially unaltered. Although the time scale of simulations is too short to show a response of the Ag deposit to changes in the Ni substrate step morphology, experiments demonstrate that the Ag deposit responds to such changes in interface morphology by gradually increasing the proportion of grains in *OR T* at the expense of those in *OR C*.

**Acknowledgments**

The authors thank Alain Ranguis (CINaM, Marseille) for his assistance with AFM measurements. D.C. and B.C. wish to thank the Agence Nationale de la Recherche for support of their research under grant ANR-GIBBS-15-CE30-0016. Substrate preparation and nickel film deposition were performed in the PLANETE cleanroom facility (CINaM, Marseille). S.S., F.R., and M.D.G. would like to acknowledge funding from a DoD Vannevar-Bush Faculty Fellowship (# N00014-16-1-2821) as well as the computational facilities of the Materials Characterization Facility at CMU under grant # MCF-677785.